\documentclass[journal, 10pt]{IEEEtran}
\IEEEoverridecommandlockouts
\usepackage{cite}
\usepackage{ifpdf}
\usepackage{stfloats}
\usepackage[tight,footnotesize]{subfigure}
\usepackage{amsmath,amssymb,amsfonts}
\usepackage{algorithmic}
\usepackage{graphicx}
\usepackage{textcomp}
\usepackage{xcolor}
\usepackage{algorithm}
\usepackage{algorithmic}
\usepackage{comment}
\def\BibTeX{{\rm B\kern-.05em{\sc i\kern-.025em b}\kern-.08em
    T\kern-.1667em\lower.7ex\hbox{E}\kern-.125emX}}

\begin{document}

\title{FlexScatter: Predictive Scheduling and Adaptive Rateless Coding for Wi-Fi Backscatter Communications in Dynamic Traffic Conditions}

\author{Xin~He,~\IEEEmembership{Member,~IEEE},
    Jingwen~Xie,
    Aohua~Zhang,
    Weiwei~Jiang,
    Yujun~Zhu, 
Tad~Matsumoto,~\IEEEmembership{Life~Fellow,~IEEE}
\thanks{This work was supported by the Natural Science Foundation of China under grant No. 62072004. Corresponding author: {\it Yujun Zhu}.}
\thanks{X. He, A. Zhang, J. Xie, and Y. Zhu are with the School of Computer and Information, Anhui Normal University, 241002, Anhui, China (e-mail: \{xin.he, zhuyujun, zhangaohua, jingwen.xie\}@ahnu.edu.cn).}
\thanks{W. Jiang is with the School of Computer Science, Nanjing University of Information Science and Technology, 210044, Jiangsu, China (e-mail: weiwei.jiang@nuist.edu.cn).}
\thanks{T. Matsumoto is with IMT Atlantique, CNRS UMR 6285, Lab-STICC, Brest, France, JAIST, and University of Oulu (e-mail: matumoto@jaist.ac.jp).}
}

\maketitle
\begin{abstract}
The potential of Wi-Fi backscatter communications systems is immense, yet challenges such as signal instability and energy constraints impose performance limits. This paper introduces FlexScatter, a Wi-Fi backscatter system using a designed scheduling strategy based on excitation prediction and rateless coding to enhance system performance. Initially, a Wi-Fi traffic prediction model is constructed by analyzing the variability of the excitation source. Then, an adaptive transmission scheduling algorithm is proposed to address the low energy consumption demands of backscatter tags, adjusting the transmission strategy according to predictive analytics and taming channel conditions. Furthermore, leveraging the benefits of low-density parity-check (LDPC) and fountain codes, a novel coding and decoding algorithm is developed, which is tailored for dynamic channel conditions.   Experimental validation shows that FlexScatter reduces bit error rates (BER) by up to 30\%, improves energy efficiency by 7\%, and increases overall system utility by 11\%, compared to conventional methods. FlexScatter’s ability to balance energy consumption and communication efficiency makes it a robust solution for future IoT applications that rely on unpredictable Wi-Fi traffic.
\end{abstract}


\section{Introduction}
Recent advancements in Internet of Things (IoT) technologies have brought backscatter communication to the forefront as a promising solution for energy-efficient data transmission, leveraging existing environmental radio frequency sources. A standard backscatter communication system consists of an excitation source, tags, and a receiver. These tags modulate and reflect signals from the excitation source to the receiver without regenerating their own signals, allowing tags to operate at ultra-low power levels, typically in the microwatt range.

Ambient backscatter communication, which utilizes ambient radio signals from devices such as Wi-Fi routers and TV transmitters, is favored over dedicated systems due to its lower deployment costs and the ubiquity of signal sources \cite{FreeRider, onay2019performance, liu2013ambient, rosenthal2019158}. The pervasive nature of Wi-Fi signals, in particular, provides substantial advantages for deployment in varied settings like shopping malls and residential areas.

However, the intermittent nature of Wi-Fi signals, governed by protocols such as IEEE 802.11a/g/n, brings significant challenges in maintaining continuous data transmission for backscatter systems~\cite{huang2010beyond}. The challenges include variable silent periods between transmissions, which can destabilize the communication link and increase packet loss of backscatter communications. Additionally, the inherent low-power characteristic of backscatter tags restricts their ability to adapt to fluctuating signal availability and interference from other environmental signals.

To address these challenges, this article introduces FlexScatter, a system designed to enhance the performance of Wi-Fi backscatter systems by leveraging predictive modeling of Wi-Fi signals. The main contributions of this paper are summarized as follows:
\begin{itemize}
    \item Adaptive backscatter scheduling: The paper introduces an adaptive transmission scheduling algorithm that dynamically adjusts transmission strategies based on real-time Wi-Fi traffic predictions. This improves energy efficiency and communication reliability in unpredictable Wi-Fi environments.
    \item 
    Wi-Fi traffic prediction: We use deep learning to predict Wi-Fi traffic patterns. These predictive models enhance the adaptability of backscatter communication, enabling dynamic adjustments to variations in Wi-Fi traffic.
    \item We integrate low-density parity-check (LDPC) codes with the principles of fountain codes, using omnipresent Wi-Fi signals for efficient and reliable data transmission. This approach adapts to varying channel conditions by adjusting the encoded packets or bit rates, ensuring optimal use of transmission resources.
\end{itemize}

The rest of this paper is organized as follows. Section II introduces related work. Section III describes the system architecture adopted in this study. In Section IV, the proposed methods are detailed. Section V presents the experimentally verified results and analysis, comprehensively evaluating the effectiveness of the proposed methods. Finally, Section VI summarizes the entire paper with concluding remarks.

\section{The State-of-the-Art}
\subsection{Backscatter Communications} 
Thanks to the significant efforts made by the research community, backscatter communications have been enabled over various radio signals. A breakthrough work in ambient backscatter communication is introduced in~\cite{liu2013ambient}, where a prototype was developed to transmit information via television signals using backscatter. LoRa backscatter~\cite{LoRaback} and PLoRa~\cite{PLoRa} enable long-range backscatter communication using LoRa signals as the excitation signal, where the frequency characteristics of the chirp signal are effectively utilized. FM backscatter~\cite{Wang17}, which uses continuous FM radio signals as the excitation, has a variety of new applications in urban areas.  

Wi-Fi-based backscatter communication has seen considerable development as well. BackFi~\cite{BackFi} operates backscatter communications over the Wi-Fi excitation signals transmitted from Wi-Fi access points with hardware modification. Wi-Fi backscatter~\cite{Bryce14} connectes the RF-powered devices to the Internet excited by a Wi-Fi signal. Passive Wi-Fi demonstrates for the first time that it can generate 802.11b backscatter transmissions using backscatter tags~\cite{Kellogg16}. HitchHike~\cite{HitchHike} enables the backscatter communication over 802.11b signals of the off-the-shelf Wi-Fi transceivers using a proposed codeword translation technique. A more recent work~\cite{Kim18} enables per-symbol and in-band backscatter communication over the Wi-Fi excitation signals using a so-called flicker detector by utilizing the residual channel knowledge of the Wi-Fi packets. Furthermore, backscattering of ultra-wideband signals is considered in~\cite{Yang17}, and X-Tandem~\cite{XTandem} enables a multi-hop backscatter system using Wi-Fi signals as excitation sources.

However, these systems have rarely considered how uncontrolled Wi-Fi traffic can be used effectively as an excitation signal. To enable more flexible backscatter communications, the inherent nature of the excitation signal must be taken into account. GuardRider~\cite{GuardRider, Yang2023} system aims at the effective utilization of the backscatter information in real Wi-Fi networks. However, it assumes that the Wi-Fi signaling duration follows Pareto distributions without verifying the correctness of the actual Wi-Fi traffic characteristics. RapidRider~\cite{RapidRider} system embeds the tag's data into a single OFDM symbol to reduce the effects of uncontrolled Wi-Fi traffic.

\vspace{-4pt}
\subsection{Overview of Traffic Prediction Techniques}
Traffic prediction in wireless networks is broadly categorized into model-driven and data-driven approaches.

\subsubsection{Model-driven approaches}
These methods model traffic as a statistical process and predict based on predefined distributions. For instance, ref.~\cite{li2017learning} treates wireless traffic as an unstable process and utilizes a statistical model based on a-stable processes for prediction. This method is particularly suitable for handling abrupt changes in self-similar traffic patterns, which are common in wireless networks.

\subsubsection{Data-driven approaches}
In contrast, data-driven approaches rely on historical traffic data for the predictions, often employing time series analysis and machine learning techniques. Ref.~\cite{xu2016big} indicates that autoregressive integrated moving average (ARIMA) models could predict regular components, but random components are complex due to low inter-correlation. Ref.~\cite{wang2019traffic} successfully predictes the arrival times of wireless network traffic using a random forest regression algorithm, outperforming traditional linear models. Refs.~\cite{trinh2018mobile} and \cite{ramakrishnan2018network} verify that long short-term memory (LSTM) and other recurrent neural networks outperform conventional statistical methods like ARIMA in predicting network traffic. Ref.~\cite{rostami2020proactive} makes further advancement of the traffic estimation by employing an LSTM-based framework to predict packet arrival times, enabling more effective dynamic scheduling in wireless networks. Hence, in this work, we implement the Wi-Fi traffic predictor using a deep-learning method. 

\section{System Architecture}

\subsection{System Overview}
Most existing Wi-Fi-based backscatter communication systems assume that the excitation source is controllable and continuous. However, to facilitate the practical deployment of the systems, it is necessary to address both the intermittent WiFi signal transmission and strong interference.

\begin{figure}[t]
\centering
\includegraphics[width=3.5in]{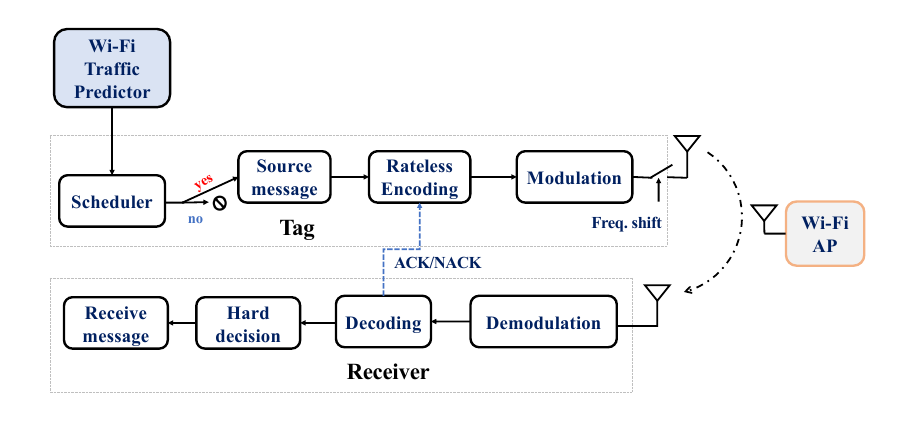}
    \caption{The proposed system model of Wi-FI backscatter communications with rateless encoding and traffic-based scheduling.}
\label{fig:fig2}
\end{figure}

As shown in Fig.~\ref{fig:fig2}, FlexScatter consists of three main components: a Wi-Fi traffic predictor, a customized tag, and a receiver. The Wi-Fi traffic predictor provides input to a scheduler in the tag, determining whether or not to transmit the source message based on the predicted traffic. If the transmission is decided, the source message is channel-encoded using a rateless coding technique, which ensures efficient and reliable data transmission even under varying channel conditions. The encoded message is then modulated and transmitted by the tag, with a $20$~MHz frequency shift. The ``roadside'' Wi-Fi access point (AP) acts as the excitation source for the backscatter communication. At the receiver, the signal is demodulated and decoded, with a hard decision block determining the final message. The system also incorporates an ACK/NACK feedback loop between the receiver and the tag, which further enhances communication reliability by allowing for a rateless coding concept. 

However, due to the ultra-low power consumption design of the tags, the Wi-Fi traffic predictor is implemented on the receiver side rather than on the tag itself. The receiver, after predicting the traffic, communicates its decision to the tag by sending a two-bit ACK signal—either `11' or `00'—to notify the tag of whether or not to proceed with the transmission. The use of a two-bit signal, instead of a one-bit signal, is crucial for improving the detection reliability at the low-power tag, ensuring that the tag can correctly interpret notifications of the receiver instructions even under challenging power constraints. This design concept enhances the overall efficiency and robustness of the FlexScatter system, particularly in scenarios where energy consumption is of crucial importance.

\subsection{Wi-Fi Traffic Predictor}
To implement the Wi-Fi traffic predictor, we employ a deep-learning-based approach leveraging a multi-scale channel attention mechanism. This algorithm leverages previously-stored traffic data to train a model capable of making accurate predictions about future Wi-Fi traffic patterns, adapting to complex and dynamic network conditions. By focusing on this approach, the system is designed to effectively balance accuracy and computational efficiency, catering to different timing scales and performance requirements.

\begin{figure*}[t]
    \centering
    \includegraphics[width=4.5in]{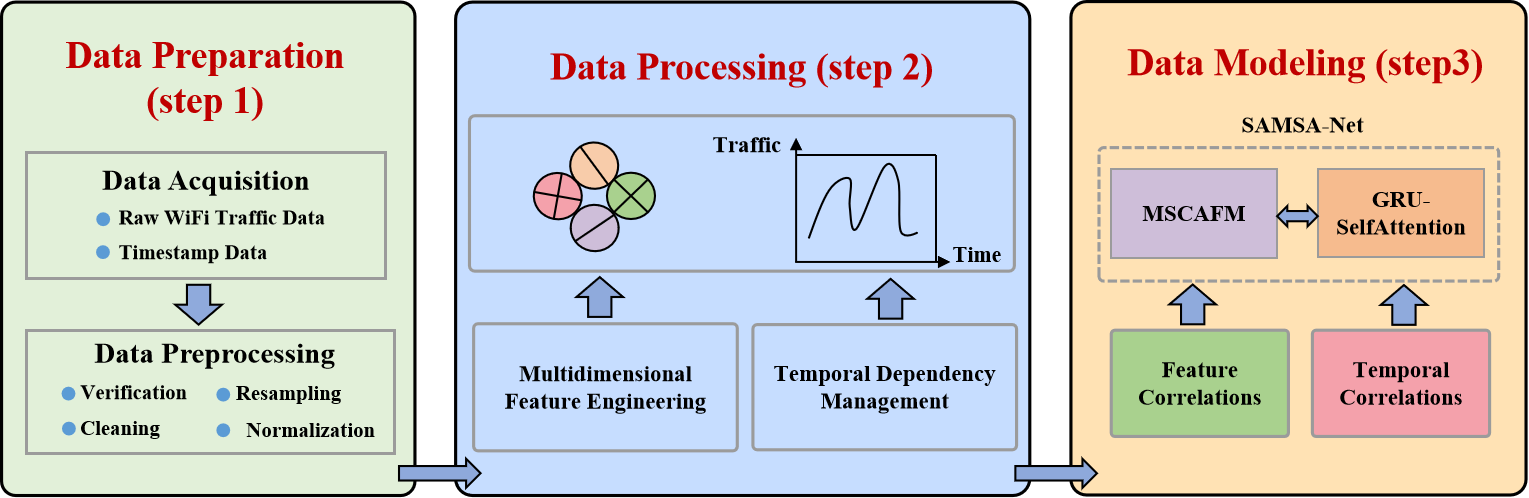}
    \caption{The workflow of the proposed method for Wi-Fi traffic prediction.}
    \label{fig:workflow}
\end{figure*}

The algorithm consists of three key stages: data preparation, data processing, and data modeling, as illustrated in Fig.~\ref{fig:workflow}. 

We collected Wi-Fi traffic data with the Edimax EW-7811Un Wi-Fi card, capturing data through the \textit{tcpdump} tool on a configured computer setup. Typically, the captured data contains the following features with different scales: time stamp, frame length (bytes), radio duration (\textmu s), interval time (s), data rate (Mbps), and signal strength (dBm). The purpose of the traffic predictor is to estimate the future interval time using these features.

Recognizing the potential for discrepancies due to device failures or malfunctions from various monitoring setups, we conduct a preprocess to refine the raw data beforehand.

\subsubsection{Data Preparation}
\begin{figure*}[t]
    \subfigure[CDF Comparison]{
        \begin{minipage}[t]{0.44\linewidth}
            \centering
            \includegraphics[width=0.95\textwidth]{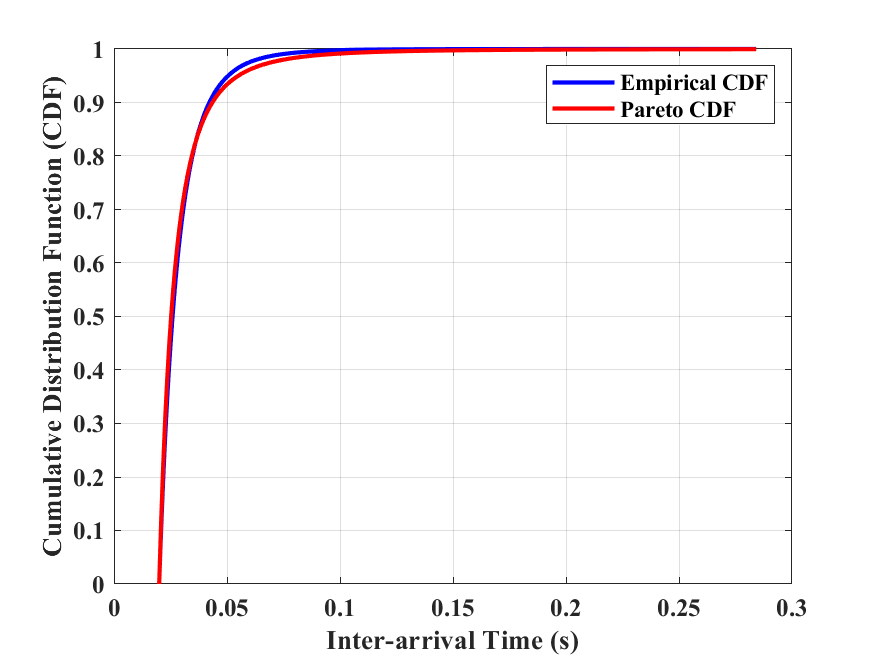}
            \label{fig: CDF Comparison}
        \end{minipage}
    }
    ~
    \subfigure[Data Distribution with]{
        \begin{minipage}[t]{0.44\linewidth}
            \centering
            \includegraphics[width=0.95\textwidth]{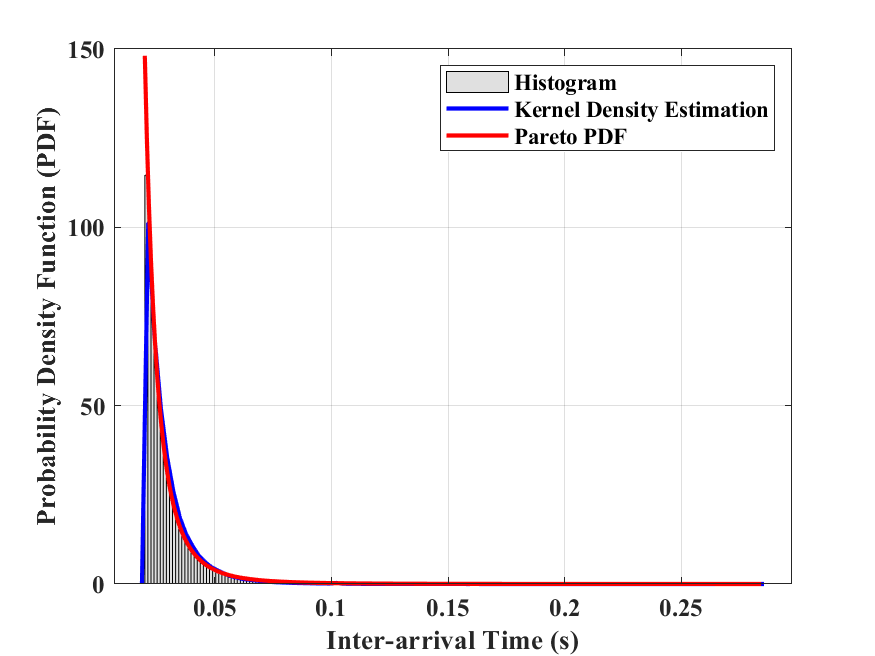}
            \label{fig: Data Distribution with}
        \end{minipage}
    } 
    \caption{Analysis of interval time data with Pareto distribution.}
    \label{fig: Analysis of interval data}
\end{figure*}

The preprocessing begins with a detailed inspection of the dataset to identify and rectify issues such as irrelevant or unquantifiable parameters, missing values, and outliers. We then slice the data into clusters using $1$~second per cluster. Particular attention is paid to the intervals between clusters of Wi-Fi data as shown in Fig.~\ref{fig: Analysis of interval data}. Following the initial cleanup, we transform the dataset to feature uniform time intervals, enhancing the consistency and analytical viability for subsequent time series evaluations. Finally, we implement a min-max normalization strategy, adjusting all feature values to a $[0, 1]$ range, which normalizes the data, facilitating more effective comparisons and analyses in the later stages.

\subsubsection{Data Processing}
The goal of Wi-Fi traffic flow prediction refers to the process of predicting the on/off state in the future based on the time series of Wi-Fi traffic data, that is, using the data of the current time \( t \) to estimate the Wi-Fi traffic at the next time \( t + \Delta t \). Furthermore, to meet the requirements of applications including Wi-Fi backscatter communications, we aggregate predictions over future time steps \(\Delta t\) to estimate overall trends. The reason is twofold; First of all, by aggregating the values within \(\Delta t\), a more accurate prediction is achieved. Then, it is extremely power-consuming for the tag to adjust the transmission frequently within a short time period. 

\subsubsection{Data Modeling}
We design a deep learning model based on the multi-scale channel attention fusion module (MSCAFM) to predict Wi-Fi traffic while considering temporal factors accurately. The model implementation comprises three key stages: multidimensional feature extraction, temporal factor fusion, and sequence prediction. 


In the multidimensional feature extraction stage, SAMSA-Net leverages MSCAFM to capture multi-scale temporal features of Wi-Fi traffic data. It combines the squeeze-and-excitation (SE) attention mechanism and gating units to optimize the representation and integration of the selected features, establishing a robust feature foundation for sequence prediction.

During the temporal factor fusion stage, the model incorporates long-term time stamps (such as hours and weekdays) as features and integrates them with traffic data through attribute feature units (AF-units). It enhances the sensitivity of the model to temporal variations, improving prediction accuracy and real-time performance.

Finally, in the sequence prediction stage, SAMSA-Net processes the integrated feature matrix using gated recurrent units (GRU) and self-attention mechanisms. It enables the model to effectively capture long- and short-term trends in traffic data, which enhances the prediction accuracy.

\subsection{Scheduler}
The designed scheduler is to enhance data transmission reliability and efficiency under varying Wi-Fi conditions. The algorithm dynamically adjusts its strategy based on the Wi-Fi traffic conditions estimated in real-time, by the Wi-Fi traffic predictor, effectively managing transmission intervals to alleviate large interference and optimizing both network reliability and energy conservation. 

Real-time Wi-Fi traffic data is collected for the current period and used to predict future traffic values for the next \(\Delta t\) seconds based on the past \(L\) time steps. In the following experiments, multiple \(\Delta t\) time step predictions are summed up to estimate overall trends and traffic levels. Experiments indicated that using \(L = 64\) and \(\Delta t = 5\) provide the best predictive performance. After that, we compare the predicted Wi-Fi interval rate with a predefined threshold $W_{I}$. If the predicted value is under $W_{I}$, then the tag will backscatter its message. Otherwise, the tag keeps silent to save energy. 

\subsection{Rateless LDPC Coding}
For the channel-encoding process, the tag implements LDPC coding with an infinite code rate, focusing on a well-designed check matrix construction and adaptive encoding and decoding processes.

\subsubsection{Index matrix construction}
LDPC codes utilize parity check matrices for encoding. To adapt to infinite code rates, we dynamically generate various index matrices by modifying initial exponent values in the index matrix, as
\begin{equation}
\mathbf{P} =
\begin{pmatrix}
a^{\mathbf{x}}b^{\mathbf{y}} & a^{\mathbf{x}+1}b^{\mathbf{y}} & \cdots & a^{\mathbf{x}+m}b^{\mathbf{y}} \\
a^{\mathbf{x}}b^{\mathbf{y}+1} & a^{\mathbf{x}+1}b^{\mathbf{y}+1} & \cdots & a^{\mathbf{x}+m}b^{\mathbf{y}+1} \\
\vdots & \vdots & \ddots & \vdots \\
a^{\mathbf{x}}b^{\mathbf{y}+n} & a^{\mathbf{x}+1}b^{\mathbf{y}+n} & \cdots & a^{\mathbf{x}+m}b^{\mathbf{y}+n}
\end{pmatrix}
\label{eq:eq8}
\end{equation}
where $a$ and $b$ stand for prime numbers, and $\mathbf{x}$ and $\mathbf{y}$ are discrete random variables characterized by their probability distributions.

\begin{figure}[t]
\centering
\includegraphics[width=3.2in]{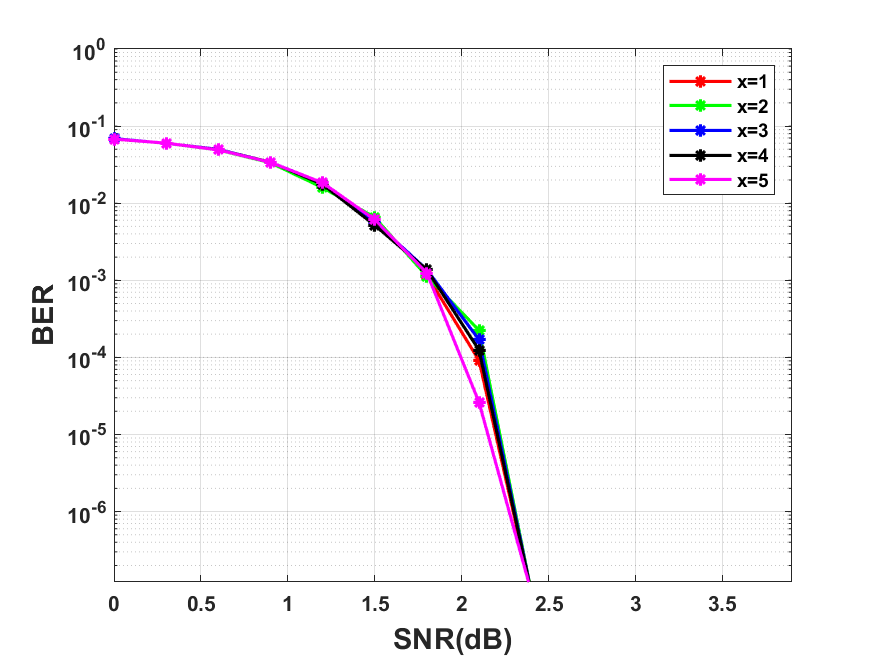}
\caption{BER performance with various initial exponent values.}
\label{fig:fig11}
\end{figure}

These index matrices facilitate encoding with different generator matrices, ensuring continuous transmission of encoded data packets. Fig.~\ref{fig:fig11} shows the simulation results demonstrating bit error rate (BER) performance with the index $x$ as a parameter. It is found that the BER performance is stable over different matrices.

\subsubsection{Efficient check matrix Design}
The system employs a new encoding technique for constructing the parity check matrix of irregular QC-LDPC codes. The check matrix \(\mathbf{H}\) consists of two parts, \(\mathbf{H}_1\) and \(\mathbf{H}_2\), where \(\mathbf{H}_1\) is derived from the index matrix, and \(\mathbf{H}_2\) includes a shiftable identity matrix \(\mathbf{I}\), facilitating the construction of a cycle-free irregular matrix. This structure not only mimics the performance of random LDPC codes but also simplifies the computation with the corresponding generator matrix.

\begin{equation}
\label{eq:eq9}
\mathbf{H}_1 = \begin{pmatrix}
\mathbf{I}_{a^xb^y} & \mathbf{I}_{a^{x+1}b^y} & \cdots & \mathbf{I}_{a^{x+m}b^y} \\
\mathbf{I}_{a^xb^{y+1}} & \mathbf{I}_{a^{x+1}b^{y+1}} & \cdots & \mathbf{I}_{a^{x+m}b^{y+1}} \\
\vdots & \vdots & \ddots & \vdots \\
\mathbf{I}_{a^xb^{y+n}} & \mathbf{I}_{a^{x+1}b^{y+n}} & \cdots & \mathbf{I}_{a^{x+m}b^{y+n}}
\end{pmatrix}
\end{equation}

\begin{equation}
\mathbf{H}_2=\begin{pmatrix}
\mathbf{I}&\mathbf{I}&0&\cdots&0&0\\
0&\mathbf{I}&\mathbf{I}&\cdots&0&0\\
\vdots&\vdots&\vdots&\ddots&\vdots&\vdots\\
\mathbf{I}&0&0&\cdots&0&0\\
0&0&0&\cdots&0&0\\
\vdots&\vdots&\vdots&\ddots&\vdots&\vdots\\
0&0&0&\cdots&\mathbf{I}&\mathbf{I}\\
\mathbf{I}&0&0&\cdots&0&\mathbf{I}
\end{pmatrix}
\label{eq:eq10}
\end{equation}

After constructing the parity check matrix \(\mathbf{H}\), it is converted into a standard form \([\mathbf{I} | \mathbf{Q}]\) through row-wise Gaussian elimination. Consequently, the generator matrix \(\mathbf{G}\) is formed as \([\mathbf{P} | \mathbf{I}]\), where \(\mathbf{P} = \mathbf{Q}^T\).

\subsubsection{Adaptive encoding and decoding of LDPC}
Unlike traditional LDPC coding, our method adopts feedback from the receiver to notify the transmitter of the encoding strategies. The tag in the system dynamically adapts encoding processes based on feedback. Positive feedback leads to successive information frame encoding, while negative feedback (NACK signals) prompts the regeneration of new index and generator matrices for re-encoding the failed frames. This adaptive encoding approach enhances the resilience of data transmission.

\subsubsection{Enhanced decoding with improved BP algorithm}
Our decoding strategy incorporates an improved Belief Propagation (BP) algorithm, which utilizes saved decoding results from previous iterations to further refine decoding accuracy in subsequent attempts. This iterative process effectively reduces BER and ensures reliable data transmission over noisy channels.

Specifically, the algorithm flow is shown in Algorithm~\ref{alg:rateless-ldpc-decoding}, where {\it BPfunction} represents the improved BP algorithm, and the normalization function represents the normalization algorithm. Let the code length of a single encoded packet be \(N\), and let \(M\) be the half of the code length. Thus, \(L(Q_i)\) represents the parity bits of the encoded packet, and \(L(C_i)\) represents the information bits of the encoded packet.

\begin{algorithm}[t]
\caption{Rateless LDPC Code Decoding Process}
\label{alg:rateless-ldpc-decoding}
\begin{algorithmic}[1]
\REQUIRE Prior probability of observations $L(C_i)$
\ENSURE Hard decision for each bit $\hat{x}$
\STATE count = 1
\WHILE {$ACK \neq 11$}
    \IF {count == 1}
        \STATE $(ACK, L(Q_i)) = \text{BPfunction}(L(C_i))$
    \ELSE
        \STATE $L(C_i) = [L(C_i)(1:M), L(Q_i)(M+1:N)]$
        \STATE Normalization function$(L(C_i))$
        \STATE $(ACK, L(Q_i)) = \text{BPfunction}(L(C_i))$
    \ENDIF
\ENDWHILE
\IF {$L(Q_i) < 0$}
    \STATE $\hat{x} = 1$
\ELSE
    \STATE $\hat{x} = 0$
\ENDIF
\end{algorithmic}
\end{algorithm}

The improved BP algorithm returns two values: a feedback signal and posterior probabilities.
\begin{itemize}
    \item The feedback signal is either ACK or NACK, represented by binary 11 and 00, respectively. ACK indicates successful decoding and informs the tag to encode and send the next information frame. Conversely, if the receiver sends a NACK, the tag will continue to use the indexing matrix module and the generator matrix module to construct more generator matrices to re-encode the unsuccessfully decoded information frame, and resend it to the receiver.
    \item The posterior probabilities consist of the probabilities of information bits and parity bits. If decoding fails, the results of the most recent information bits are retained and used to replace the information bits in the next BP algorithm iteration, while the parity bits in the new packet remain unchanged. Therefore, in practical system operation, apart from the first transmission, which requires a complete encoded packet, subsequent transmissions only need to send encoded parity bits if decoding fails.
\end{itemize}

The normalization algorithm ensures that the results after each BP algorithm iteration are within a certain range to maintain successful decoding. It is particularly important because a high signal-to-noise ratio (SNR) can affect the posterior probability range in the improved BP algorithm, potentially generating erroneous values and leading to decoding failures. In the experiment detailed in the following Section, the normalization algorithm employs a linear function to limit the posterior probability values within the range of -5 to 5.

\begin{figure}[t]
\centering
\includegraphics[width=3.2in]{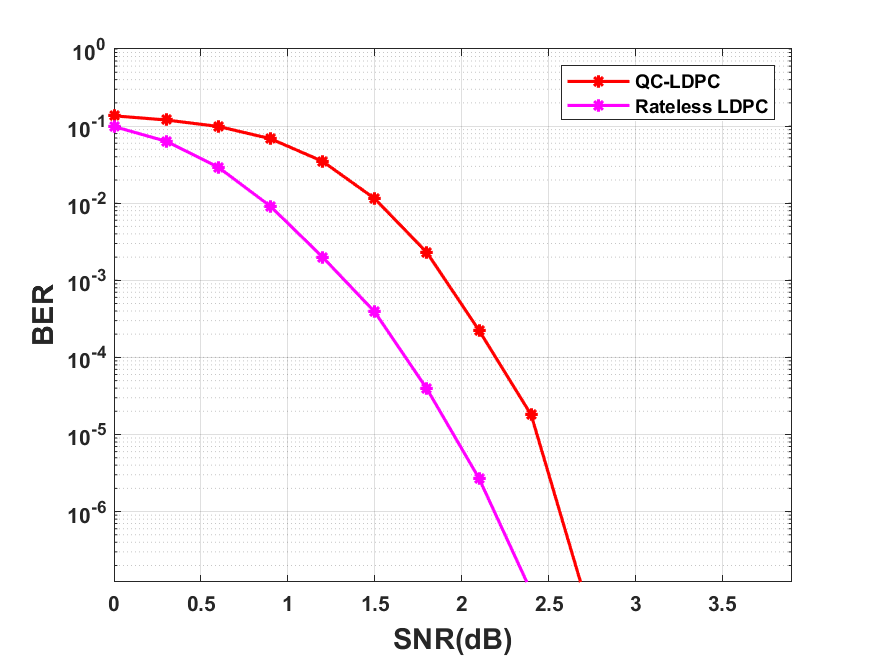}
\caption{BER comparison between rateless and QC-LDPC codes across different SNRs.}
\label{fig:fig16}
\end{figure}

\begin{figure}[t]
\centering
\includegraphics[width=3.2in]{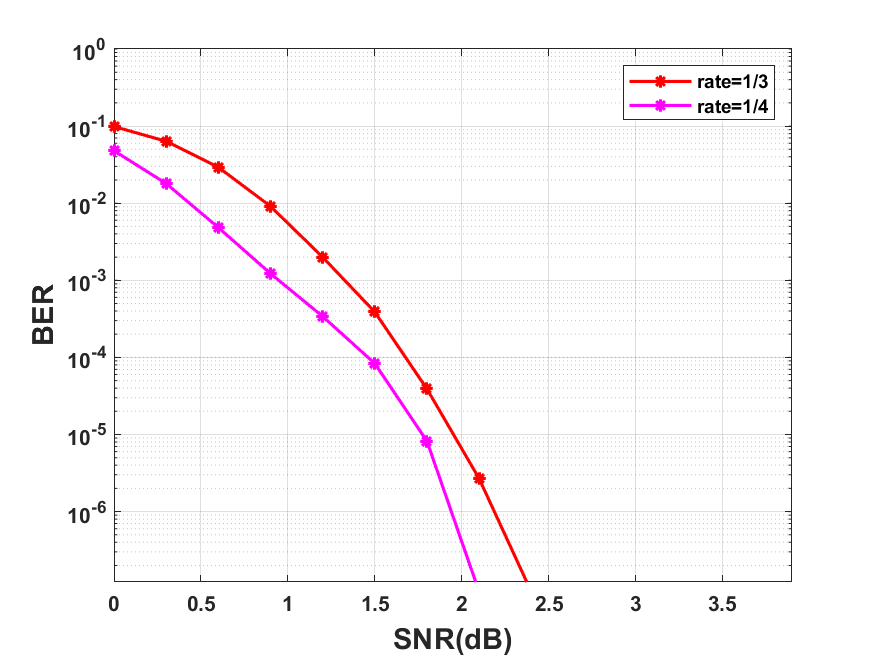}
\caption{Effect of adjusting the coding rate on BER of rateless LDPC codes.}
\label{fig:fig17}
\end{figure}

\begin{figure}[t]
\centering
\includegraphics[width=3.2in]{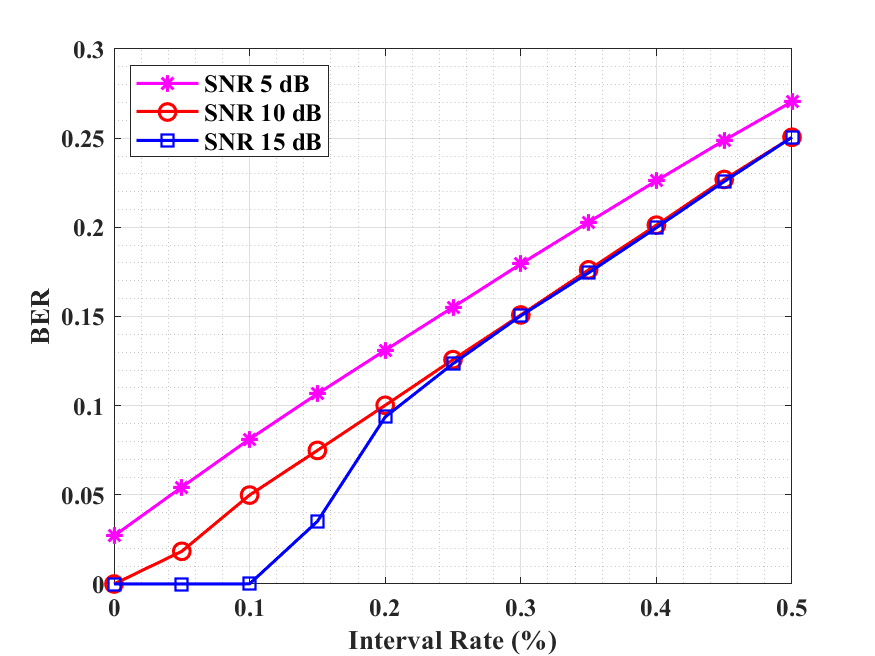}
\caption{Performance comparison of different SNR across varied Wi-Fi traffic interval ratios.}
\label{fig:SNR0.5_vs_Interval}
\end{figure}

\section{Experimental Results and Analysis}

\subsection{Performance Metrics}
Typically, we consider the BER indicating the transmission accuracy. Besides, energy consumption is a crucial factor for the tag requiring ultra-low power consumption. Hence, we jointly consider BER, throughput, and energy consumption when designing a new metric, namely, utility in the backscatter communications. 

For evaluating energy efficiency, first of all, based on the ultra-low power consumption characteristics of backscatter communication systems, this paper assumes that each bit transmitted by the backscatter tag consumes 10 \textmu J of energy. According to Algorithm~\ref{alg:rateless-ldpc-decoding}, except for the first transmission, which requires a complete coded packet, subsequent transmissions only need to send parity bits. Therefore, the total energy required to send a coded packet is given by
\begin{equation}
E = 10 \left( \frac{N}{2}(n+1) \right)
\label{eq:energy},
\end{equation}
where $n$ represents the total number of packets and $N$ the number of bits in each coded packet. 

To describe system performance in terms of how much information is correctly received per unit of time, we adopt throughput as an indicator of reliable transmission performance, defined as:
\begin{equation}
T = R (1 - P_e)
\label{eq:throughput},
\end{equation}
where $R$ is the transmission rate at the tag and $P_e$ the BER after $n$ transmissions.

With the energy and throughput metrics, the utility value \(U\) is defined to evaluate system performance:
\begin{equation}
U = \frac{\alpha T}{\beta E}
\label{eq:utility}.
\end{equation}
In this work, \(\alpha\) and \(\beta\) are set at 1 for a general evaluation.

To make solid evaluations of the system performance, we consider three distinct cases: 
\begin{itemize}
    \item Without scheduling: In this scenario, the system operates without any predictive scheduling, relying solely on the immediate availability of Wi-Fi signals, which may result in suboptimal performance due to the unpredictability of traffic.

    \item Scheduling using deep learning: Here, the system employs the proposed deep-learning-based algorithm for traffic prediction, enabling it to schedule transmissions more effectively by anticipating Wi-Fi traffic patterns. 

    \item Scheduling using ideal values: In this case, the system uses perfect, idealistic predictions of Wi-Fi traffic (i.e., knowing the exact future traffic conditions). This serves as an upper-bound benchmark, showing the maximum possible performance improvement that could be achieved with perfect prediction.
\end{itemize}

\subsection{Micro-Benchmark Experiments}
\subsubsection{Evaluation on coding/decoding design}
This section evaluates the performance of rateless LDPC codes with different communication scenarios, focusing on BER as the primary performance metric. The experiments are conducted on an additive white Gaussian noise (AWGN) channel to simulate varying noise levels represented by different SNRs.

The simulation transmits 1000 packets in total, with 1310 bits per packet. Each packet is encoded using LDPC with rate $1/2$ and the index matrix parameters being \(a=3\) and \(b=7\).

Initial experiments aim to compare the rateless LDPC codes with conventional QC-LDPC codes. As shown in Fig.~\ref{fig:fig16}, the rateless codes consistently outperformed QC-LDPC codes in terms of BER over the entire SNR values tested, demonstrating their robustness in noisy environments.

The simulation then changed the coding rate to evaluate the BER performance sensitivity to the coding rate matching. The results reveal that an increase in the number of transmitted packets notably decreased BER, highlighting the superiority of rateless codes to maintain low error rates, as shown in Fig.~\ref{fig:fig17}.

Due to the interval rate of Wi-Fi traffic affecting the performance largely, we adopt a set of Wi-Fi interval ratios to evaluate the performance of rateless LDPC codes. As demonstrated in Fig.~\ref{fig:SNR0.5_vs_Interval}, the adopted rateless LDPC code could still perform effectively until 20\% interval ratios with different SNR values. Hence, we set the threshold of the interval ratio to 20\% in the scheduling algorithm. 

\begin{figure}[t]
\centering
\includegraphics[width=3.2in]{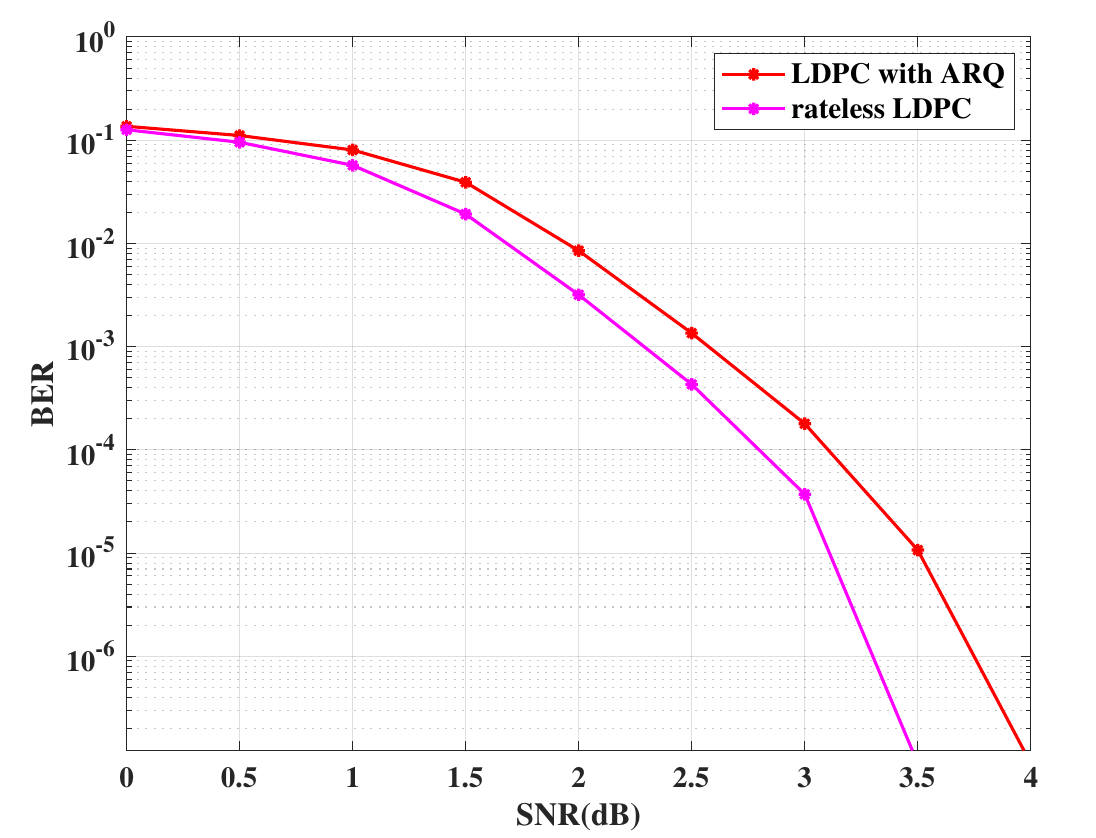}
\caption{Performance comparison of rateless LDPC coding and ARQ under varied Wi-Fi conditions.}
\label{fig:ARQ_vs_rateless}
\end{figure}

Lastly, a comparative evaluation against a basic ARQ retransmission mechanism underlined the superior adaptability of rateless LDPC codes to fluctuating network conditions, as shown in Fig.~\ref{fig:ARQ_vs_rateless}. In the ARQ scheme, the construction matrix of the LDPC code remains the same and we just send the same coded packets repeatedly. Compared to the ARQ case, we gain around $0.5$ dB in SNR. 

These extensive tests validate the superior performance of rateless LDPC codes, making them highly suitable for environments with variable interference and discontinuity of Wi-Fi traffic.

\subsubsection{Wi-Fi traffics prediction} 
The Wi-Fi traffic dataset adopted in this study was captured in commercial shopping venues, covering the main operational hours from 9:00 AM to 10:00 PM daily over a week. An 80/20 data split strategy was employed to identify the best model configuration.

A sliding window technique was employed, utilizing the data at the current time \( t \) to predict the Wi-Fi traffic at the next timing \( t + \Delta t \). The sliding window approach is chosen because it allows the model to capture temporal dependencies in the data, making it particularly effective for time-series prediction. By clustering similar patterns within the window, the model can accurately predict and anticipate future traffic trends.

In the experiments, the model hyperparameters included a learning rate of 0.0001, a batch size of 64, and a training period of 100 epochs. We utilize the Adam optimizer for gradient descent, and all neural network-based methods were implemented using the PyTorch framework due to its flexibility and convenience for deep learning research.

To evaluate the performance of our prediction models, we select three primary performance metrics: mean squared error (MSE), root mean squared error (RMSE), and mean absolute error (MAE).

To demonstrate the superiority and effectiveness of our proposed method, we compare it with several widely used baseline models, including ARIMA, MLP, CNN, LSTM, GRU, and CNN-GRU. The results are summarized in Table~\ref{tab:experimental_results}.

\begin{table}[t]
\centering
\caption{Comparison of Experimental Results under Different Settings}
\label{tab:experimental_results}
\begin{tabular}{c|c|c|c}
\hline
Method       & MSE     & RMSE    & MAE     \\ \hline
ARIMA        & 0.027819 & 0.166790 & 0.103130 \\ \hline
MLP          & 0.012733 & 0.112840 & 0.082369 \\ \hline
CNN          & 0.011440 & 0.106960 & 0.084562 \\ \hline
LSTM         & 0.009120 & 0.095499 & 0.070014 \\ \hline
GRU          & 0.008823 & 0.093932 & 0.062518 \\ \hline
CNN-GRU      & 0.008659 & 0.093056 & 0.063019 \\ \hline
Our Method   & \textbf{0.008285} & \textbf{0.091021} & \textbf{0.061926} \\ \hline
\end{tabular}
\end{table}

The experimental results clearly demonstrate that our deep learning methods significantly outperform the traditional time-series models such as MLP and ARIMA in terms of prediction accuracy for the next \(\Delta t = 5\). It is found that our proposed method can achieve substantial improvements in MSE, RMSE, and MAE compared to these models. Specifically, our method reduces MSE by approximately 70.22\%, RMSE by 45.43\%, and MAE by 39.95\% compared to the ARIMA model; it reduces MSE by approximately 34.93\%, RMSE by 19.34\%, and MAE by 24.82\% compared to the MLP model.

Moreover, when compared with other deep learning models such as CNN, LSTM, GRU, and CNN-GRU, our method also demonstrates superior performance. Our method reduces MSE, RMSE, and MAE by 27.58\%, 14.90\%, and 26.77\% compared to CNN; 9.16\%, 4.69\%, and 11.55\% compared to LSTM; 6.10\%, 3.10\%, and 0.95\% compared to GRU; and 4.32\%, 2.19\%, and 1.73\% compared to CNN-GRU. These results not only asure the effectiveness of our method in the field of deep learning but also highlight its significant advantages in time-series prediction.

\begin{figure}[t]
    \centering
    \includegraphics[width=3.25in]{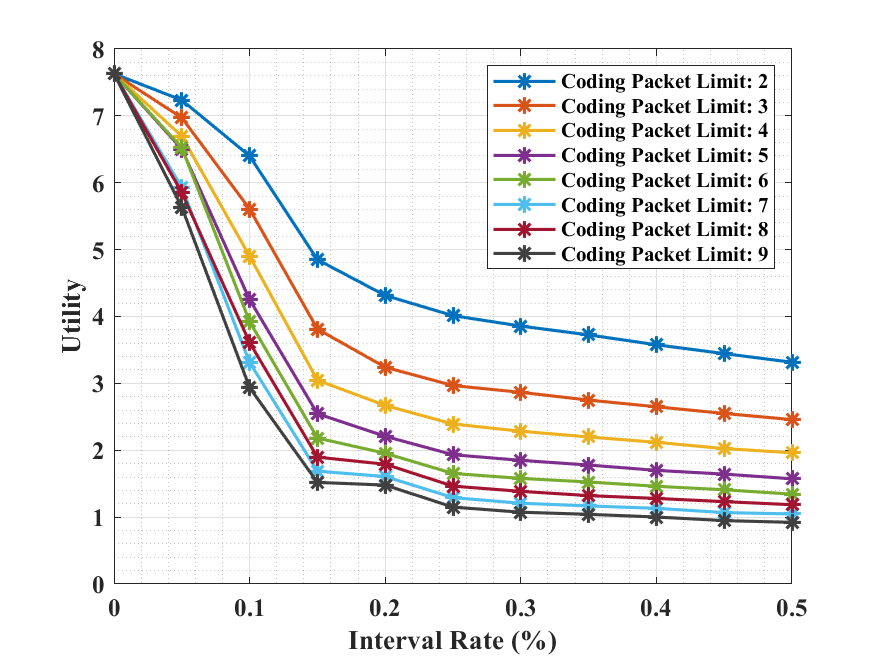}
    \caption{Comparison of System Utility Values at Different Thresholds.}
    \label{fig:figThresholds}
\end{figure}

\subsubsection{Impact of Scheduling Threshold}
To optimize the scheduling strategy for backscatter tags, we propose halting data transmission when the Wi-Fi interval rate exceeds a predefined threshold to conserve energy. Backscatter tags will resume transmission only when the interval rate is below this threshold. During low interval rates, the number of encoding packets is restricted based on fountain coding principles to efficiently utilize limited energy resources.

We then investigate the utility value variations under different coding rates as the interval rate increases, with a fixed SNR of 10 dB in an AWGN channel. The Wi-Fi interval rate ranges from 0 to 0.5, using a rateless LDPC coding scheme. The maximum number of encoding packets was randomly changed between 2 and 9, and the data transmission rate $R$ was set at $1$ Mbps. The utility value was used as the primary performance indicator of our design system.

Figure~\ref{fig:figThresholds} presents utility values against Wi-Fi interval rates with maximum encoding packet limit as a parameter. The utility value curves plateau between interval rates of 15\% to 25\%. The plateau occurs because the total energy used remains constant as the number of encoding packets needed for reliable transmission increases, leading to stabilization in throughput and utility value.

Increasing the maximum number of encoding packets results in the stable utility value with the same interval rate. This is because each additional packet monotonically increases the transmission energy, while throughput increase diminishes, leading to stable utility value, as shown in \eqref{eq:energy}.

To avoid low utility value regions, it is recommended that the interval rate threshold $W_I$ be set at 25\%. Backscatter tags should remain silent when the Wi-Fi interval rate exceeds this threshold and resume transmission when it falls below the thereshold. The optimal coding rate is 1/5, and the maximum number of encoding packets should be set to 4, with which optimizing the transmission utility.

\subsection{Macro-benchmark Experiments}
This section analyzes the impacts of scheduling strategies using deep learning predictive methods on the performance enhancement of backscatter communication systems. Our simulations were conducted in realistic environments characterized by AWGN with an SNR of 10 dB. The scheduling strategies used in the simulations are a 0.25-time slot interval rate threshold and a 1/5 code rate, and are benchmarked against systems without scheduling strategies under the same conditions.

\begin{table}[t]
\centering
\caption{Pareto Distribution Parameters for Different Scenarios}
\begin{tabular}{c|c|c}
\hline
\textbf{Scenario} & \textbf{Channel} & $\boldsymbol{\alpha}$ \\
\hline
           & 1 & 0.57 \\
Laboratory & 6 & 0.54 \\
           & 11 & 0.43 \\
\hline
              & 1 & 0.57 \\
Shopping Mall & 6 & 0.54 \\
              & 11 & 0.43 \\
\hline
                      & 1 & 0.16 \\
Residential Apartment & 6 & 0.09 \\
                      & 11 & 0.04 \\
\hline
\end{tabular}
\label{tab:pareto_parameters}
\end{table}

\begin{figure*}[t]
    \centering
    \subfigure[BER]{
        \begin{minipage}[t]{0.16\linewidth}
            \centering
            \includegraphics[width=0.95\textwidth]{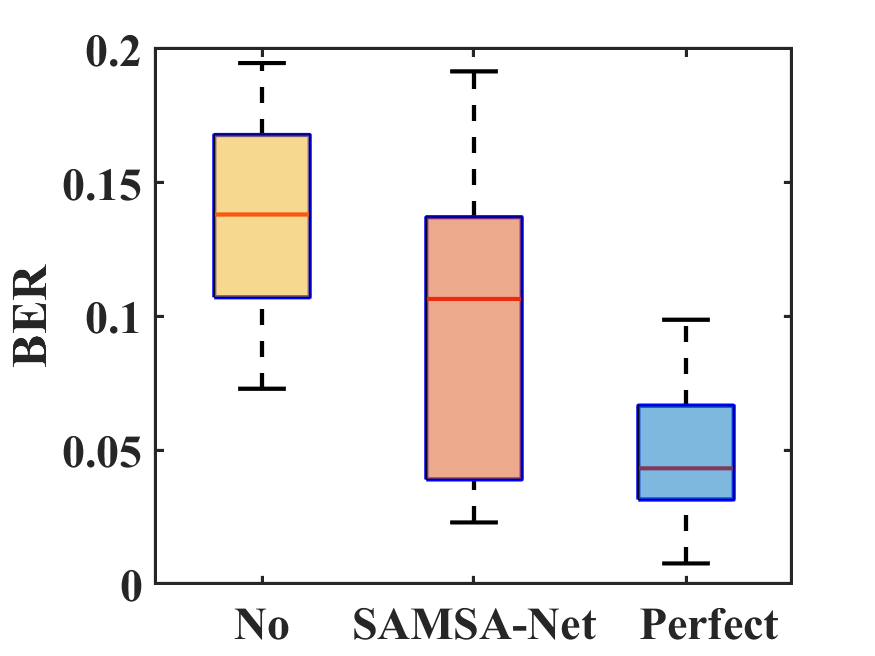}
            \label{fig:BER}
        \end{minipage}
    }
    \subfigure[energy]{
        \begin{minipage}[t]{0.16\linewidth}
            \centering
            \includegraphics[width=0.95\textwidth]{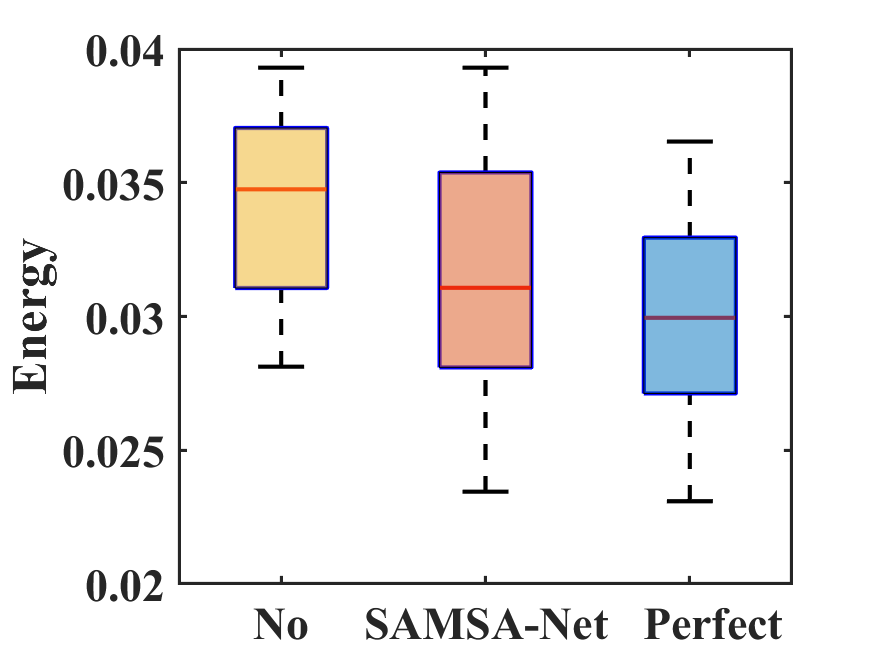}
            \label{fig:Energy}
        \end{minipage}
    }
     \subfigure[utility]{
        \begin{minipage}[t]{0.16\linewidth}
            \centering
            \includegraphics[width=0.95\textwidth]{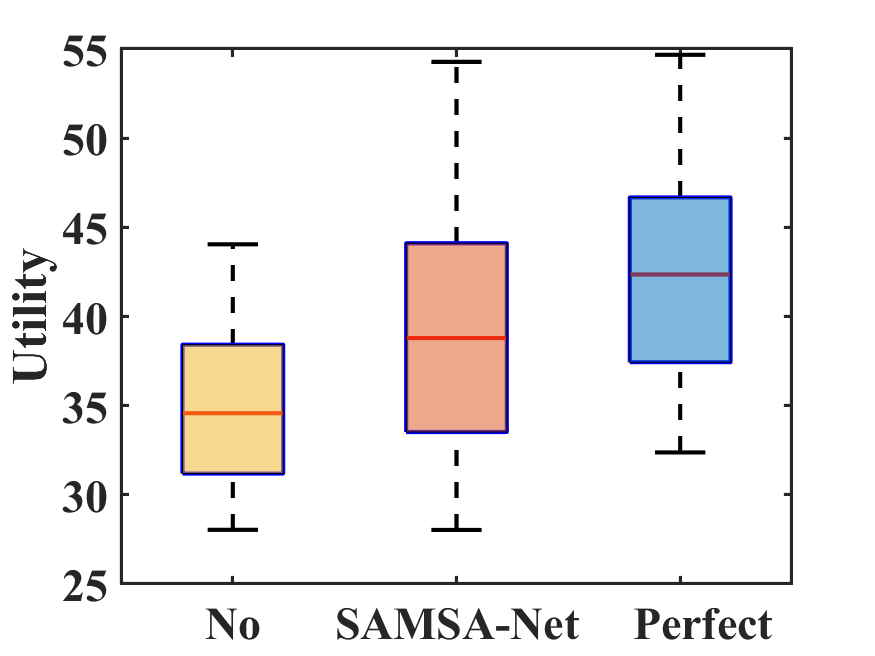}
            \label{fig:Utility}
        \end{minipage}
    }
    \subfigure[throughput]{
        \begin{minipage}[t]{0.16\linewidth}
            \centering
            \includegraphics[width=0.95\textwidth]{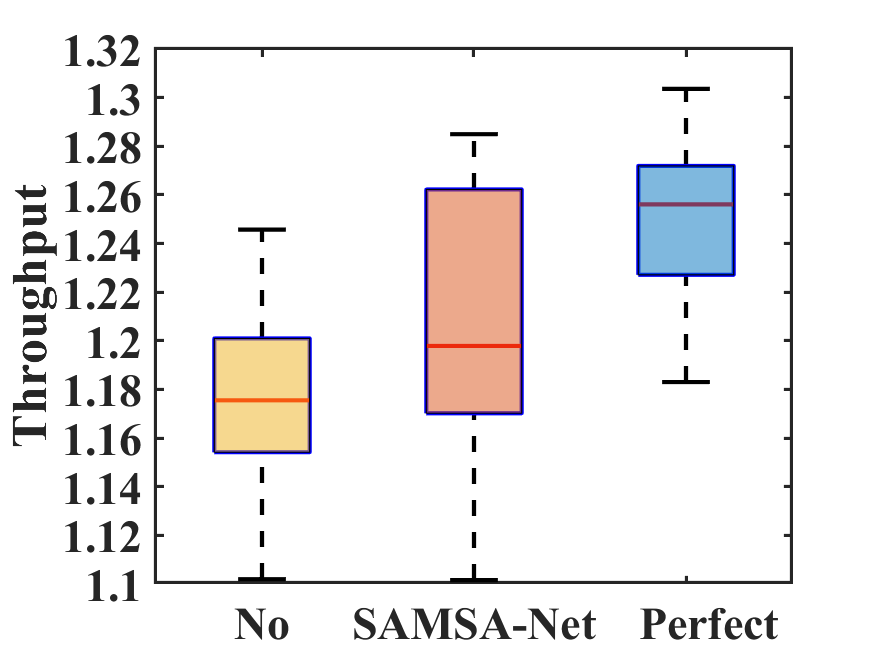}
            \label{fig:5f}
        \end{minipage}
    }
    \subfigure[time]{
        \begin{minipage}[t]{0.16\linewidth}
            \centering
            \includegraphics[width=0.95\textwidth]{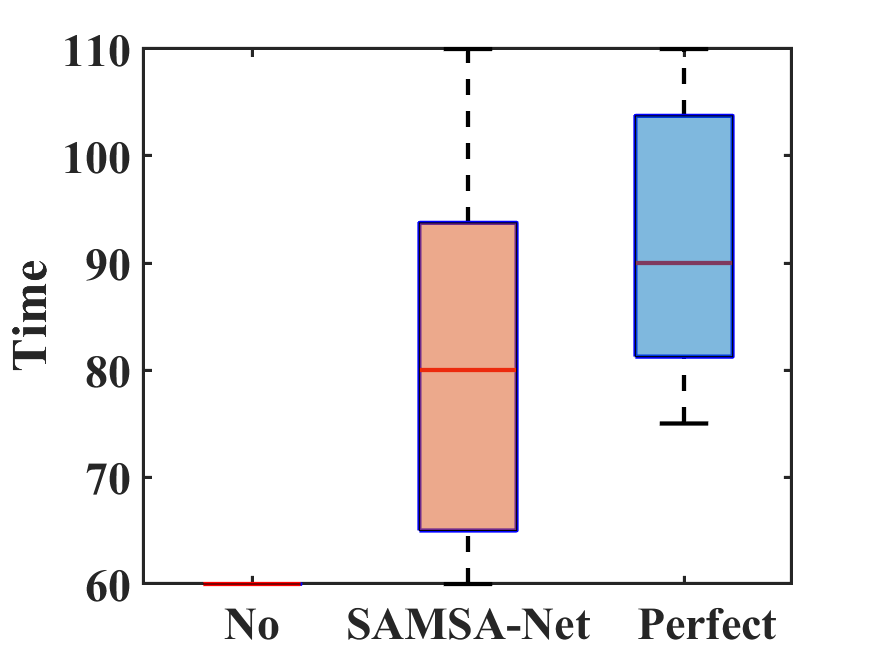}
            \label{fig:Time}
        \end{minipage}
    }
    \caption{The improvement on the system performance using scheduling driven from deep learning and perfect knowledge. The perfect knowledge is adopted as the optimal performance for performance comparison.}
    \label{fig:system_metrics1}
\end{figure*}

\subsubsection{Impact of Scheduling}
To test the system performance of the scheduling and rateless coding as a whole, we randomly selected 20 time slices from the test set of the captured Wi-Fi data. Each time slice contains a total number of 24,000 data frames to send. 

As shown in Fig.~\ref{fig:system_metrics1}, it is found that, specifically, the BER decreases by approximately 29.7\%; transmission energy by approximately 6.8\% in average. Throughput increases by approximately 2.5\% in average. The utility factor increases by approximately 11.1\% in average. The results indicate that using deep learning models for predictive scheduling significantly enhances the accuracy and efficiency of system's transmission performance.

Idealistic scheduling conditions, serving as a benchmark for the system's optimal performance, result in an average decrease of 64.9\% in BER and 12.5\% in transmission energy, while throughput and utility increase by 6.1\% and 21.6\%, respectively, compared to no scheduling case. These results highlight the superiority of scheduling algorithms under optimal conditions.

Through comparative analysis, although predictive scheduling can not achieve the optimal results of idealistic scheduling, it still shows significant advantages over no scheduling conditions. Specifically, predictive scheduling exhibits noticeable decreases in key performance metrics such as BER, and transmission energy, while throughput and utility metrics also show significant improvements. This indicates that despite limitations in predictive accuracy, the overall performance of predictive scheduling outperforms the case without scheduling and approaches the best performance under the idealistic scheduling conditions, validating the effectiveness and rationality of the predictive scheduling strategy. Although transmission time is slightly extended, it remains within an acceptable range, ensuring a balance between transmission efficiency and system reliability.

In summary, the experimental results demonstrate that the predictive scheduling strategy significantly improves system performance and efficiency compared to no scheduling. These findings not only validate the effectiveness of our proposed predictive scheduling strategy not only in enhancing transmission accuracy and system throughput but also in assuring its rationality and potential in practical applications.
\begin{figure*}[t]
    \subfigure[BER]{
        \begin{minipage}[t]{0.3\linewidth}
            \centering
            \includegraphics[width=0.95\textwidth]{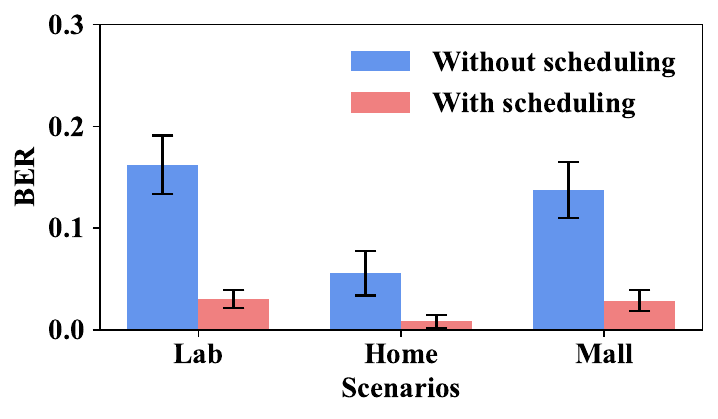}
            \label{fig:ber_three_scenarios}
        \end{minipage}
    }
    ~
     \subfigure[utility]{
        \begin{minipage}[t]{0.3\linewidth}
            \centering
            \includegraphics[width=0.95\textwidth]{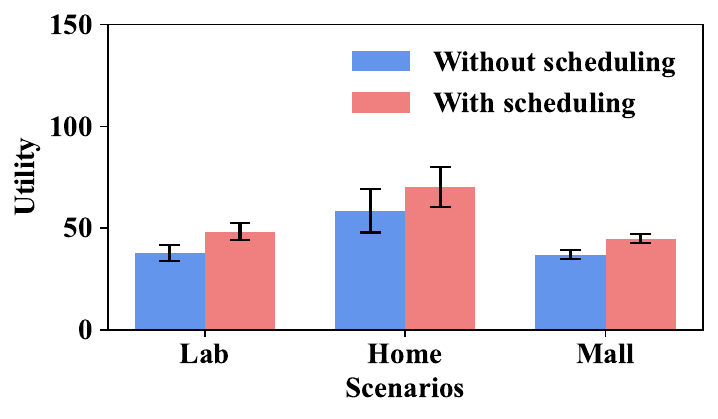}
            \label{fig:utility_three_scenarios}
        \end{minipage}
    }
    ~
\subfigure[throughput]{
        \begin{minipage}[t]{0.3\linewidth}
            \centering
            \includegraphics[width=0.95\textwidth]{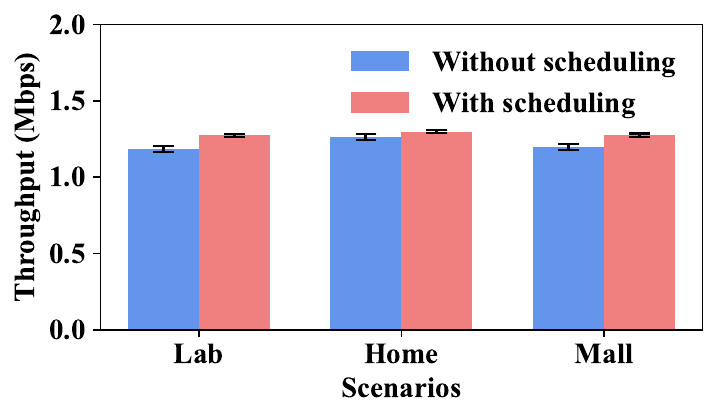}
            \label{fig:througuput_three_scenarios}
        \end{minipage}
    }
    \caption{The performance with/without scheduling in three different scenarios using re-generated packets using Pareto distribution with the obtained parameters from the captured data.}
    \label{fig:performance_three_scenarios}
\end{figure*}
\subsubsection{Impact of Sites} 
After analyzing the Wi-Fi traffic data in three different scenarios: a shopping mall, a laboratory, and a residential apartment, it is found that the cumulative distribution function of the idle state of the excitation source fits the Pareto distribution, as shown in Fig.~\ref{fig: Analysis of interval data}. Based on the collected Wi-Fi data, we estimate the Pareto distribution parameters $\alpha$ in these three scenarios, as shown in Table~\ref{tab:pareto_parameters}, where $\alpha$ represents the shape parameter of the Pareto distribution; the larger the $\alpha$, the thinner the tail of the distribution, indicating a lower probability of large idle times.

To demonstrate the scheduling strategy's effectiveness, we generate packet intervals using the estimated Pareto distribution parameters with which we conducted simulations having 500 runs. Each simulation uses 5 milliseconds as a time slice, transmitting 100 frames per time slice.

As shown in Fig.~\ref{fig:performance_three_scenarios}, the scheduling strategy significantly improves performance metrics compared to the non-scheduling strategy in all three scenarios. 

In the laboratory scenario, scheduling reduces the BER by 80\%, from 0.1607 to 0.033, a fivefold improvement. Throughput increases by 7\% (from 1.1869 to 1.2743), and utility improves by 22\% (from 38.7053 to 47.1117).

In the residential scenario, scheduling reduces the average BER by 83\%, from 0.0559 to 0.0092, a sixfold improvement. Throughput rises by 3\% (from 1.2626 to 1.2981), and utility improves by 21\% (from 59.7128 to 72.2397).

In the shopping mall scenario, scheduling reduces the average BER by 81\%, from 0.1467 to 0.0278, a fivefold improvement. Throughput increases by 7\% (from 1.1968 to 1.2774), and utility improves by 24\% (from 36.4366 to 45.1348).

Through the detailed evaluations with the three scenarios, it can be concluded that the scheduling strategy significantly reduces BER in the popular environments exemplified by the three scenarios, demonstrating its substantial effectiveness. Initial BER levels vary, with the residential scenario exhibiting a lower initial BER compared to the laboratory and shopping mall scenarios. The scheduling strategy shows particularly expected effects in the laboratory and shopping mall scenarios, highlighting its benefits in high-traffic and complex environments. Moreover, the scheduling strategy enhances throughput in all three scenarios, improving both transmission quality and energy efficiency. Finally, it increases the system utility ratio in all scenarios, with more substantial advantages in high-traffic and complex environments.

The performance improvements indicate that our scheduling strategy is practical and applicable to diverse environments. It enhances system reliability and efficiency, which are supported by theoretical and empirical verifications for real-world deployments.

\section{Conclusion}
In this paper, we have proposed a novel deep learning-based traffic prediction
and coding technique, FlexScatter, for Wi-Fi backscatter communications with 
uncontrolled traffics. The effectiveness and practicality of the coding and scheduling technique proposed in this paper have been verified through simulations with the aim of its application to the Wi-Fi backscatter communication systems. By leveraging deep learning-based traffic prediction, we have designed a scheduling algorithm which effectively identifies whether to transmit or to stay silent. Furthermore, we have proposed a rateless LDPC code to tackle the problem of dynamically varying channel conditions. It has been shown that the proposed system significantly enhances both reliability and efficiency. The improvements in terms of reliability and efficiency have been demonstrated not only in controlled laboratory environments but also in real-world application scenarios, to provide strong support for the deployment in diverse and complex settings.

\bibliographystyle{IEEEtran}
\bibliography{reference.bib}

\end{document}